\documentclass[]{spie}  %>>> use for US letter paper
\usepackage{amsmath,amsfonts,amssymb}
\usepackage{graphicx}
\usepackage[colorlinks=true, allcolors=blue]{hyperref}
\usepackage[numbers]{natbib}

\title{X-ray Hybrid CMOS Detectors:
Recent Development and Characterization Progress} %\\

\author[a]{Tanmoy Chattopadhyay}
\author[a]{Abraham D. Falcone}
\author[a]{David N. Burrows}
\author[a]{Samuel Hull}
\author[a]{Evan Bray}
\author[a]{Mitchell Wages}
\author[a]{Maria Macquaide}
\author[a]{Lazar Buntic}
\author[a]{Ryan Crum}
\author[a]{Jessica O'Dell}
\author[a]{Tyler Anderson}

\affil[a]{Pennsylvania State University, Department of Astronomy \& Astrophysics, University Park, PA 16802, USA}

\authorinfo{Further author information: (Send correspondence to T. Chattopadhyay)\\T.C.: E-mail: txc344@psu.edu, Telephone: 1 814 852 9733\\}

\begin{document} 
\maketitle

\begin{abstract}
X-ray Hybrid CMOS Detectors (HCDs) have advantages 
over X-ray CCDs due to their higher readout rate abilities, flexible readout, inherent radiation hardness, and low power, which make them more suitable for the 
next generation large-area X-ray telescope missions. 
The Penn State high energy astronomy laboratory has been working on the 
development and characterization of HCDs 
in collaboration with Teledyne Imaging Sensors (TIS). A custom-made H2RG 
detector with 36 $\mu$m pixel pitch and 18 $\mu$m ROIC shows an improved 
performance over standard H1RG detectors, primarily due to a reduced level 
of inter-pixel capacitance crosstalk (IPC). However, the energy resolution 
and the noise of the detector and readout system  are still limited when 
utilizing a SIDECAR at non-cryogenic temperatures. We characterized an 
H2RG detector with a Cryo-SIDECAR readout and controller, and we find an 
improved energy resolution of $\sim$2.7 \% at 5.9 keV and read noise of 
$\sim$6.5 e-. Detections of the $\sim$0.525 keV Oxygen K$\alpha$ and $\sim$0.277 keV Carbon K$\alpha$ lines with this detector display an improved sensitivity level at lower energies. This detector was 
successfully flown on NASA's first water recovery sounding rocket flight 
on April 4$^{th}$, 2018. We have also been 
developing several new HCDs with potential applications for future X-ray 
astronomy missions. We are characterizing the performance of small-pixel 
HCDs (12.5 $\mu$m pitch), which are important for the development of a 
next-generation high-resolution imager with HCDs. The latest results on these small pixel detectors has shown them to have the best read noise and energy resolution to-date for any X-ray HCD, with a measured 5.5 e- read noise for a detector with in-pixel correlated double sampling. Event recognition in HCDs 
is another exciting prospect. We characterized a 64 $\times$ 64 pixel 
prototype Speedster-EXD detector that uses comparators in each pixel to 
read out only those pixels having detectable signal, thereby providing an 
order of magnitude improvement in the effective readout rate. Currently, 
we are working on the development of a large area Speedster-EXD with a 
550 $\times$ 550 pixel array. HCDs can also be 
utilized as a large FOV instrument to study the prompt and afterglow 
emissions of GRBs and detect black hole transients. In this context, 
we are characterizing a Lobster-HCD system for future CubeSat experiments. 
This paper briefly presents these new developments and 
experimental results.
\end{abstract}

% Include a list of keywords after the abstract 

\section{INTRODUCTION}
\label{sec:intro}  % \label{} allows reference to this section

X-ray Charge Coupled Devices (CCDs) \citep{Lesser15_ccd,gruner02_ccd} 
have been the work-horse for soft X-ray instrumentation
for more than two decades now, having been implemented in several X-ray 
astronomy missions, e.g., ASCA, Chandra, XMM-Newton, Swift, Integral, 
Hitomi, and AstroSat.
These detectors provide moderate energy resolution, very low electronic
read noise, and small pixel sizes that enable high angular resolution images of astronomical sources for Chandra-like X-ray mirrors.  
Existing missions have been extremely 
successful, and the next generation missions will build on that success by combining fine angular resolution and large collecting area in order to probe deeper into the high redshift and low luminosity universe \citep{vikhlinin12_smartx}. Lynx (previously
known as X-Ray Surveyor) \citep{gaskin15_lynx}, NASA's 2020 decadal survey mission, for 
example, plans to have 30 times higher collecting area than Chandra. 
These upcoming missions, therefore, require focal plane detectors with 
faster readout speed than the existing modern-day X-ray CCDs, in order to avoid pile up and saturation effects in the detectors \citep{lumb00_pileup_xmm}. 

The new generation soft X-ray detectors should fulfill the following 
requirements imposed by upcoming large telescope missions (e.g. Lynx). 
\begin{itemize}
\item The detectors should provide fast readout to avoid pile up effects. 
\item The detectors should provide nearly Fano-limited energy resolution with low electronic 
read noise for quality spectroscopic data.
\item The pixel size should be small to provide high angular
resolution
images. 
\item The detectors should be radiation hard which is important for
the detectors to operate longer and provide quality data over the years.
\end{itemize}   
X-ray Hybrid CMOS detectors (hereafter X-ray HCDs) \citep{bai08} are active pixel sensors
and are expected to fulfill all of these requirements. The other advantage 
of these detectors is the low power requirement compared to CCDs (at least
by a factor of 10). The Pennsylvania State University (PSU) has been 
collaborating with Teledyne Imaging Sensors (TIS) for more than a decade
now on the technological development of X-ray HCDs and their 
characterization. PSU is involved in multiple activities and projects
on X-ray HCDs to continue its technological development and 
improvement in the measurements for future astronomical missions \citep{hull17}.

In Sec. \ref{hcd}, we describe the main characteristics of X-ray 
HCDs followed up by  
recent improvement in energy resolution and read noise measurements for 
an H2RG detector with cryogenic SIDECAR in Sec. \ref{hxrg}. 
The same H2RG detector is 
successfully launched in NASA's first water recovery rocket experiment
in early 2018. We will briefly describe the camera on this payload and the calibration of 
the instrument in Sec. \ref{wrx}. Sec. \ref{other} briefly describes the 
new developments in X-ray HCDs, including both small-pixel detectors with in-pixel CDS and HCDs with event-driven readout and on-chip digitization, followed up by new experiments with 
X-ray HCDs to demonstrate their
radiation hardness, sub-pixel localization capability, and wide field 
imaging capability.

\section{X-ray hybrid cmos detectors (HCDs)}\label{hcd}

HCDs are active pixel sensors. For more than a decade now, PSU has been collaborating with TIS to develop X-ray HCDs
based on TIS HAWAII series detectors \citep{loose03_h2rg}. X-ray HCDs are composed of two separate
layers: one absorbing Silicon layer to absorb the photons and convert the 
deposited energy into charge, and the other Silicon layer is for the 
Readout Integrated Circuit (ROIC) to read out the charge of the 
individual pixels and convert them into voltage signal (see Fig. \ref{hcd}). 
\begin{figure} [ht]
   \begin{center}
   \includegraphics[scale=0.28]{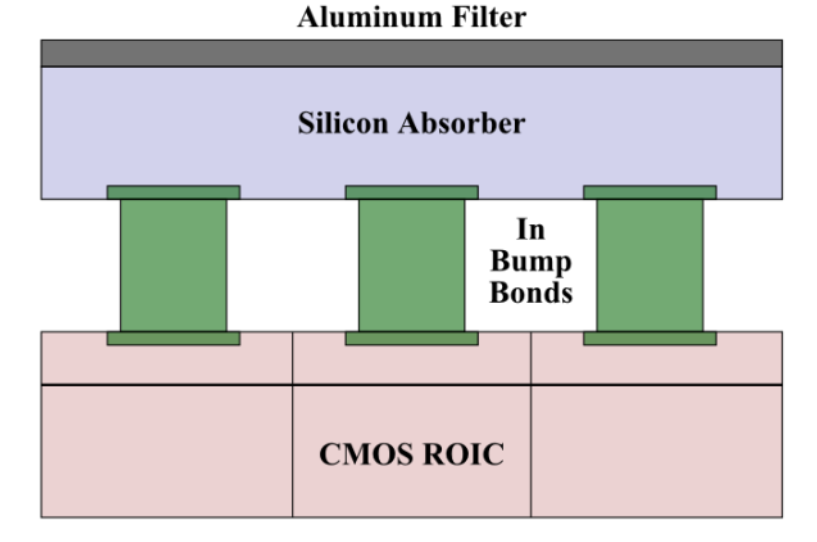}
\includegraphics[scale=0.5]{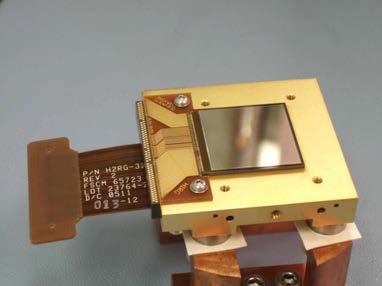}\\
   \end{center}
   \caption
{Left: Cross-sectional schematic of hybrid CMOS X-ray detector. Si absorbing layer is bump bonded to the read out
electronics layer in each pixel, allowing for separate optimization.
Right: Image of an H1RG hybrid CMOS detector (`1' stands for 1024 $\times$ 1024 pixel array) with 18 $\mu$m pixel pitch. There is a 50 nm Al layer 
deposited directly on the Si layer.
}
\label{hcd} 
\end{figure}
The pixels in the 
Silicon layer are attached to the ROIC pixels by Indium bump bonds.
Because of this kind of architecture of separate layers for absorption
of photons and electronics, each layer can be optimized separately. 
For example, the absorbing layer can be made highly resistive so that
the depletion depth can be as high as 500 $\mu$m making these detectors
sensitive to high energies $>$ 20 keV. On the other hand, the readout layer
can also be optimized to accommodate new electronic components without 
affecting the Silicon absorbing layer.

HCDs provide several advantages over the modern day CCDs as described below.
\begin{itemize}
\item \textit{Fast readout:} HCDs typically provide an order of magnitude faster readout speed compared to CCDs. 
The SIDECAR also facilitates parallel readout (4/16/32) channels with 
readout speeds of $>$ 10 Mpixel/s for each channel. High readout speed ensures the probability of 
pile up in the detector pixels to be extremely small, which is important 
in the context of large telescope area instruments. HCDs also provide readout
in window mode where only a certain region of the detector plane will be 
readout. This feature is useful while observing bright transient events since this window can be made arbitrarily small in order to achieve huge sub-frame rates.
\item \textit{Radiation hardness:} CCDs are susceptible to proton 
displacement damage, and there are well known detector performance effects from radiation
damage of the Chandra CCDs; e.g. \citep{grant05_radiation_damage_chandra}. Unlike transferring 
the charge across the pixels as in case of CCDs, HCDs readout the 
individual pixels. Therefore even if a HCD pixel is damaged, the whole column 
is not affected by that damage as is the case for CCDs.

Unlike CCDs, the gates of HCDs are not directly exposed, which makes HCDs
less susceptible to micro-meteoroid damage \citep{struder01_mircrometeoroid_damage_xmm}.
\item \textit{Low power:} HCDs are low power devices. Amount of power 
consumed by an HCD is of the order of mW. For comparison, the CCD in XRT Swift \citep{burrows05_swift} consumes $\sim$8 watts of power. An HCD can do all such operations
for $\sim$100 $-$ 200 mW of power. This is an important feature of HCDs which make 
them suitable for future X-ray astronomy missions.
\end{itemize}   
 
\section {Characterization of HxRG detectors}\label{hxrg}

The first generation of the X-ray HCD devices delivered by TIS to PSU in 
2006 were Hybrid Visible Silicon (HyViSi) detectors with H1RG (`1' stands
for 1024 $\times$ 1024 pixel array) CMOS ROIC in which we demonstrated the
successful replacement of of the normal anti-reflective coating (used for optical/IR detectors) with a 
500 nm optical blocking filter directly deposited on the silicon 
detection layer \citep{falcone07}. A Teledyne SIDECAR 
\footnote{http://www.teledyne-si.com/ps-sidecar-asic.html} is used to provide 
the clock and bias signals to the HxRGs. It also provides chip programming, 
signal amplification, analog-to-digital conversion, and data buffering. 
The SIDECAR is followed up by SIDECAR Acquisition Module (SAM) which provides 
further signal processing and amplification.

The modified H1RG detectors with 18 $\mu$m pixel pitch (see Fig.
\ref{hcd}) were characterized
in detail using a PSU cube stand, which is a light tight vacuum chamber
with pressure $\sim$10$^{-6}$ torr. The details of the experiment set up
and the results have been reported in \citep{falcone07,falcone12,bongiorno09,prieskorn13,prieskorn14_qe,hull17}. 
The best energy resolution and read noise for H1RG detectors 
were found to be 4.2 \% at 5.9 keV and 16 e- respectively.
The read noise and energy resolution are found to be limited by 
\begin{enumerate}
\item Inter-pixel capacitance cross-talk (IPC) due to change in gate voltage 
while integrating the charge degrades the energy resolution. 
We experimentally estimated the IPC to be
around 10 \% of the signal shared with a given neighboring pixel. 
\item There can be thermal voltage fluctuation in the SIDECAR ASIC during the integration of the charge. This effect can be 
reduced significantly by cooling the ASIC below 200 K. 
\item Pixel to pixel gain variation also contribute to the read noise and 
energy resolution of the detector. 
\end{enumerate}   

In order to reduce IPC (as described in [1]), a 2048 $\times$
2048 pixel H2RG readout integrated circuit (ROIC) was bump bonded to 
a 36 $\mu$m pitch Silicon detector layer (with only a subset of 
the ROIC pixels bonded), such that the effective pixel
size is 36 $\mu$m instead of 18 $\mu$m. This reduced the IPC 
significantly from 10 \% to 1.8 \% \citep{prieskorn13}. 
It is to be noted 
that in the modern X-ray HCD devices, implementation of 
capacitive transimpedance amplifier (CTIA) eliminates IPC, 
which will be discussed in later sections.

\subsection{Characterization of an H2RG with cryogenic 
SIDECAR$^{\rm TM}$}

In order to investigate the performance of the same H2RG detector with SIDECAR
when cooled to cryogenic temperatures, we utilized a cryogenic
SIDECAR ASIC from TIS. Cooling the ASIC to temperatures $<$ 200 K, with minimal thermal fluctuations, is expected to provide improved detector performance.

\begin{figure} [ht]
   \begin{center}
   \includegraphics[scale=0.23]{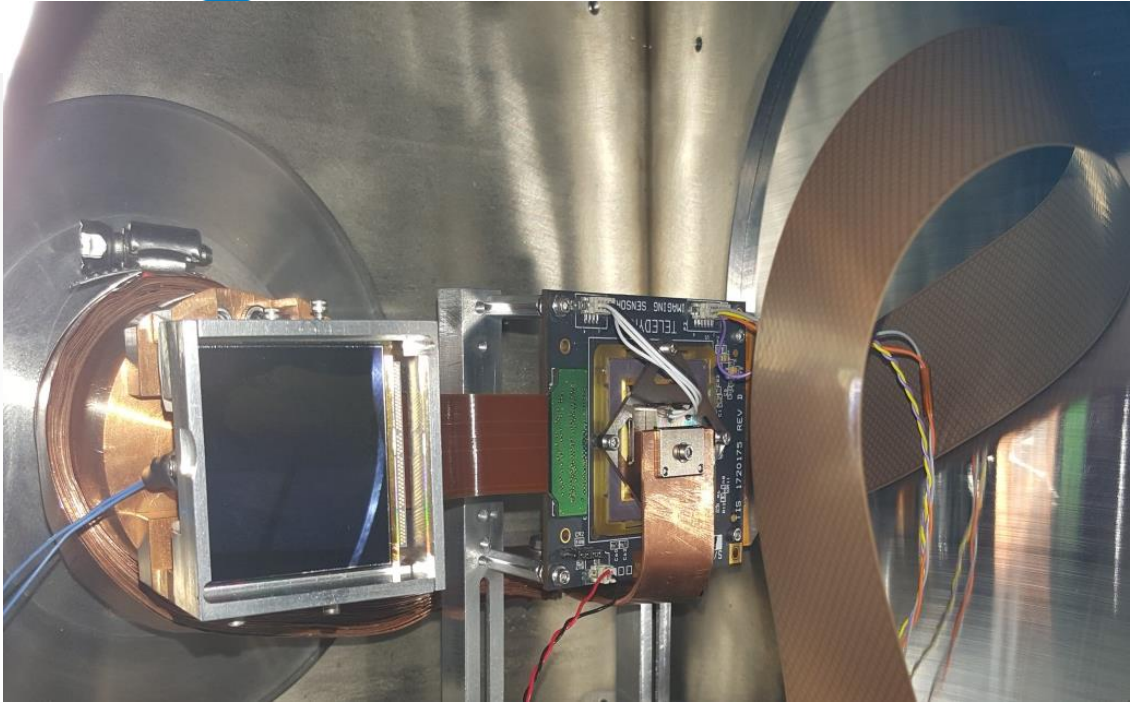}
   \end{center}
   \caption
{The interior of the test chamber with an H2RG hybrid CMOS detector 
attached to a cold finger and the cryogenic SIDECAR$^{TM}$ to the right 
of the detector, with a cold strap to cool the SIDECAR$^{TM}$.
}
\label{h2rg_cryo}
\end{figure}

The experiment set up is shown in Fig. \ref{h2rg_cryo}. 
The H2RG is directly 
mounted on the cold finger, whereas a cold strap is connected between
the cold finger and the ASIC. The detector temperature is kept at 130 K 
using a Labview-controlled PID loop. The SIDECAR ASIC temperature was at
180 K. The detector was run in two different modes $-$ 32 channel and 4 channel readout with frame times of 1.49 seconds and 10.65 seconds respectively. The 
raw images are processed using standard IDL-based 
routines where we first determine the channel to channel
gain and offset variation and apply those corrections to CDS images. This is 
followed up by offset correction for each row in the image. 

\begin{figure} [ht]
   \begin{center}
\includegraphics[scale=0.42]{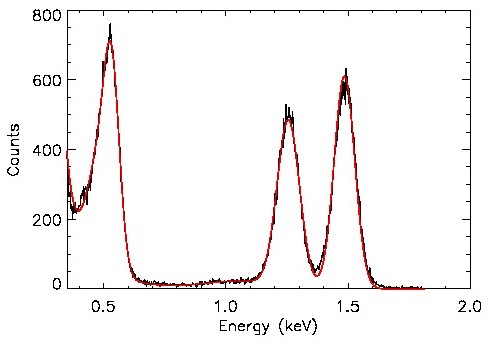}
\includegraphics[scale=0.42]{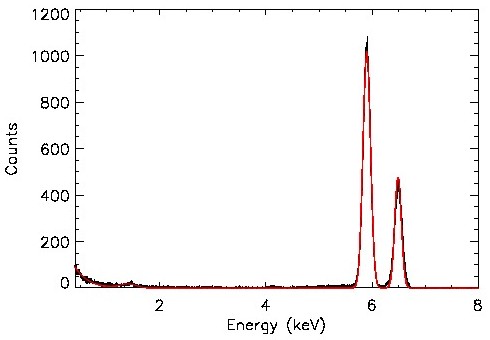}
 \end{center}
   \caption
{Left: An H2RG spectrum of O K$\alpha$ (0.525 keV), Mg K$\alpha$ 
(1.25 keV), and Al K$\alpha$ (1.49 keV) 
lines. A fit to the data is shown in red. Right: Same for Mn K$\alpha$ 
and K$\beta$ lines at 5.9 kev and 6.4 keV.
}
\label{h2rg_spec}
\end{figure}

For characterization of the detector, we used an Fe$^{55}$ radioactive source
which emits 5.9 keV and 6.4 keV photons, as well as a Po$^{210}$ source to induce 
fluorescence from targets that provided several low energy X-ray lines ranging from 0.525 keV (O K$\alpha$) to 4.5 keV (Ti K$\alpha$). 
We used single pixel events for the analysis. The spectra from these measurements are shown in 
Fig. \ref{h2rg_spec}. 
The Oxygen line is detected and well resolved, due partly to the improvement in the performance of 
the H2RG detector with the use of Cryo-SIDECAR operated at low temperature. %It implies that the
%low energy threshold of the device is around 0.3 keV. 
It is to be noted that energy resolution of 2.7 \% at 5.9 keV 
is the best ever energy resolution reported
for an X-ray H2RG detector. We estimate the read noise to be $\sim$6.5 e-.  

\section{X-ray HCD in WRX rocket payload}\label{wrx}
The same H2RG detector with a Cryo-SIDECAR was recently flown in
the Water Recovery X-ray (WRX) sounding rocket payload 
that was successfully launched from Kwajalein
Atoll on April 4$^{th}$, 2018. Notably, it was the first NASA astrophysics 
sounding rocket payload to attempt water
recovery, and at the same time this is first time an X-ray HCD was 
flown in space. The payload consists of primary optics followed by  
off-plane reflection gratings \citep{mcentaffer13_grating}, and a camera with the H2RG detector along with the cryo-SIDECAR and a custom camera interface board. A wide field of view (3.25$^\circ$ $\times$ 3.25$^\circ$) enabled observations of a large region of the Vela supernova remnant, with a science goal to observe 3rd and 4th order 
OXII. Key technology goals were the technology demonstration in a space 
environment of X-ray hybrid CMOS detectors, and the demonstration of the water recovery techniques and technology.
The technology readiness level (TRL) of X-ray HCDs was successfully raised to TRL 9. Analysis of the flight data is currently under progress and will be 
reported elsewhere. In this section, we briefly describe the instrument and the ground calibration of the payload. 

The spectrometer design consists of a mechanical collimator, X-ray 
reflection gratings, a mirror module, and the hybrid CMOS detector camera. 
%Fig. \ref{wrx} shows a schematic diagram of 
%the full instrument light path.
%\begin{figure} [ht]
%   \begin{center}
%   \includegraphics[scale=0.25]{schematic_wrx.png}
% \end{center}
%   \caption
%{Schematic of WRXR spectrometer. X-rays enter the instrument through the 
%collimator on the left before undergoing diffraction in the grating array, 
%and then reflecting onto the detector over 2 meters away.
%}
%\label{wrx}
%\end{figure}
X-rays entering through the lightweight wire-grid collimator 
are intercepted by the off-plane
reflection gratings and dispersed to produce spectral
lines on the detector plane with $\sim$190 mm cross-dispersion extent. 
Because the size of the H2RG detector is 35 mm $\times$ 35 mm, this would
result in less than 20 \% of a given spectral line falling on the 
detector area. To mitigate this large line dispersion, a mirror module is inserted between the reflection gratings and the camera. This module contains an array 
of nickel-coated mirrors that reflect light
back onto the detector area to minimize photon losses from the gratings.
For a more complete description of the instrument design and full 
specifications, see \citep{miles17}.

\begin{figure} [ht]
   \begin{center}
   \includegraphics[scale=0.085]{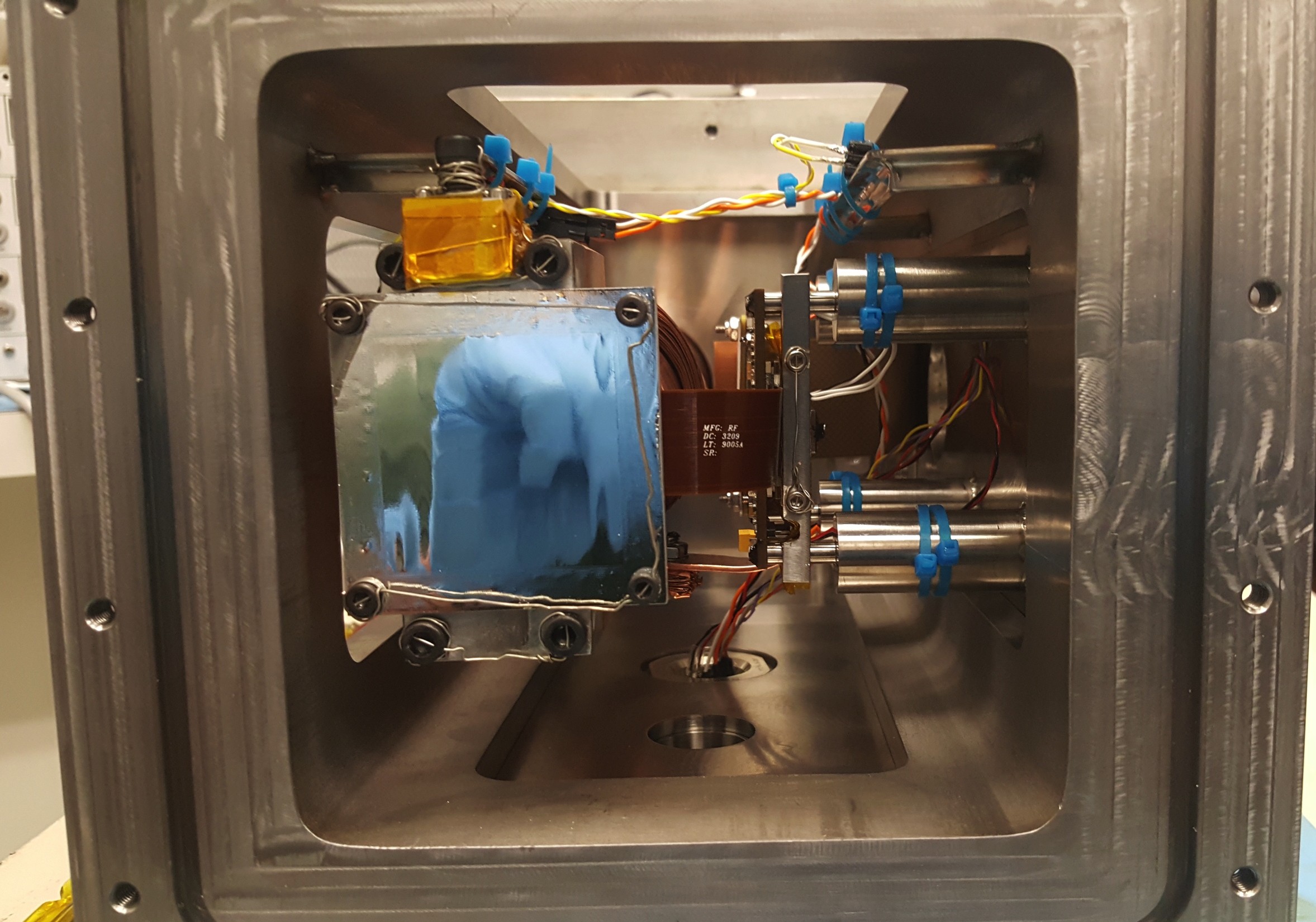}
\includegraphics[scale=0.085]{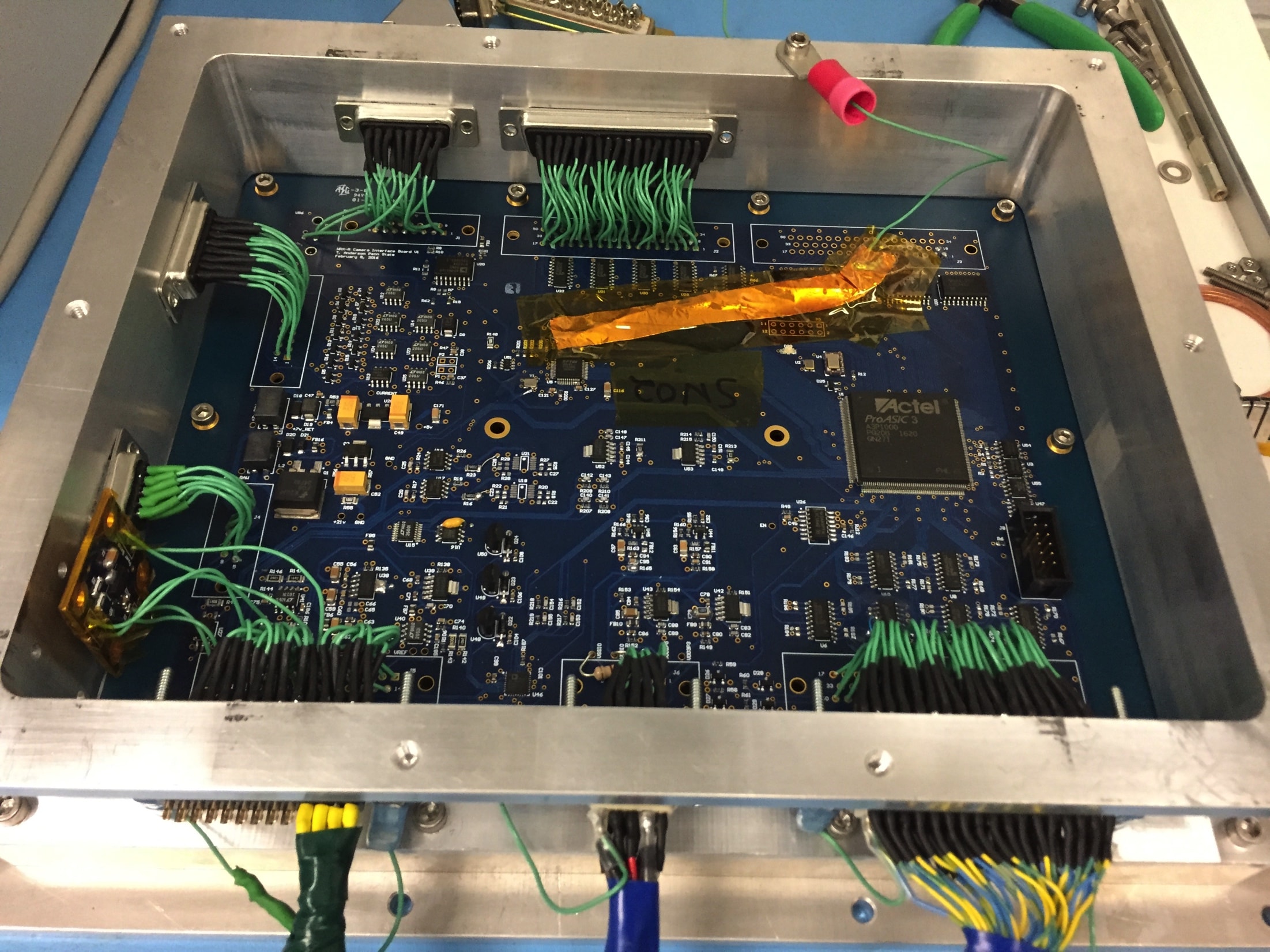}
 \end{center}
   \caption
{Left: Picture of the interior of the camera chamber. The H2RG detector is 
mounted on the cold finger. The reflection of the blue glove can be 
seen on the optical filter. The cryo-SIDECAR to the right of the detector 
is mounted on the side wall using four standoffs. There is a Fe$^{55}$
calibration source at the top left corner of the filter.
Right: Image of the Camera Interface Board (CIB) developed at PSU. The 
CIB interfaces the Cryo-SIDECAR and the spacecraft. The CIB is attached
to the back wall of the chamber.
}
\label{h2rg_wrx}
\end{figure}

The camera package is a custom built 8.5 $\times$ 8.5 $\times$ 12.75 inch 
enclosure that is mounted at the spectrometer
focal plane (see Fig. \ref{h2rg_wrx} (left)). Inside, a copper cold finger 
cools the H2RG and the cryogenic SIDECAR to 130 K and 175 K respectively. This temperature was cold-biased to a temperature that is much lower than required since the camera was expected to warm up during launch and flight, after the liquid nitrogen was pulled away.  
Because this particular H2RG detector does not have an aluminum filter deposited 
directly onto it, a 45 nm Ti plus 70 nm Al filter was
procured from Luxel and installed in front of the detector surface. A
Fe$^{55}$ calibration source was mounted inside the enclosure 
to enable valuable in-flight 
calibration X-rays (seen above the filter in Fig. \ref{h2rg_wrx}). 
The detector housing was
isolated from the main instrument vacuum section by a controllable 
GN2 actuated gate valve, protecting
the detector from external factors when not pointed at the science 
target; an on-board ion pump was used to maintain the
vacuum pressure in the enclosure. 
At PSU, we developed a Camera Interface Board (CIB) which interacts with 
the SIDECAR ASIC to provide power, filtering, and data buffering
(shown in the right panel in Fig. \ref{h2rg_wrx}). The successful use of this CIB board in space also raises the CIB to TRL-9. 
The detector was run with 32 parallel output channels, 
providing a frame time of 1.48 seconds.
A 15 V bias voltage was applied to the H2RG substrate. 

\begin{figure} [ht]
   \begin{center}
   \includegraphics[scale=0.35]{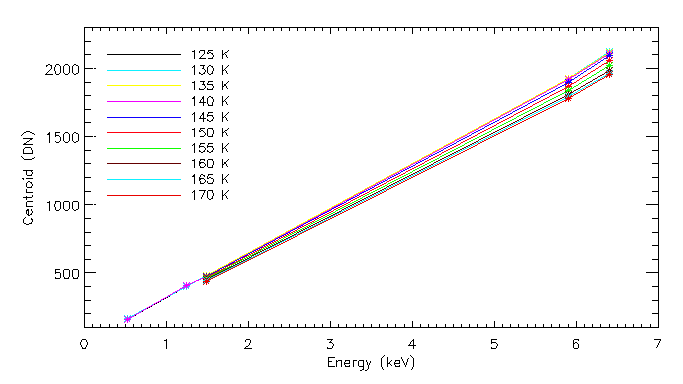}
\includegraphics[scale=0.35]{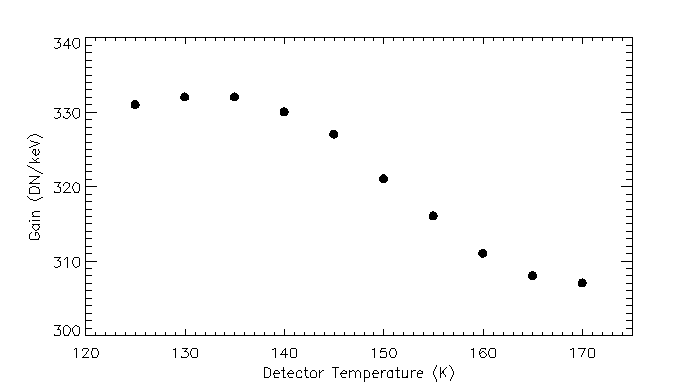}\\
\includegraphics[scale=0.35]{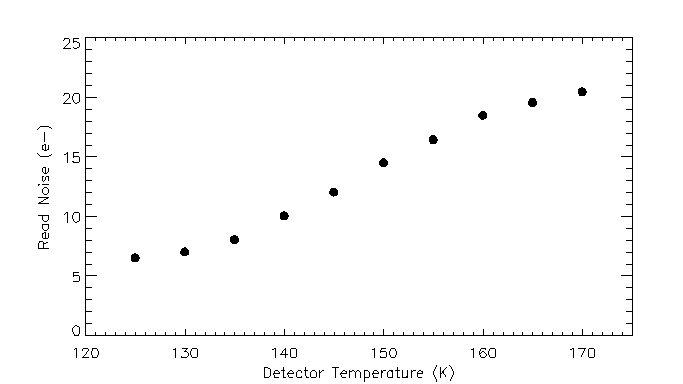}
\includegraphics[scale=0.35]{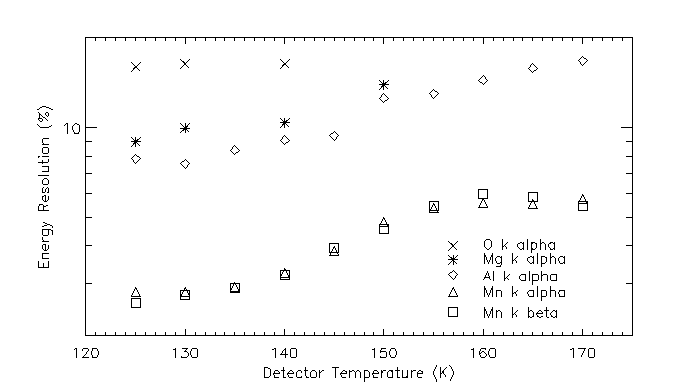}
 \end{center}
   \caption
{Top-left: Linearity of the detector in the energy range of 
0.525 $-$ 6.4 keV at multiple temperatures from 125 K to 170 K. 
Top-right: The gain of the detector as a function of temperature. 
Bottom:left: Dependence of read noise of the camera on
temperature. 
Bottom:right: Change in energy resolution of the camera with temperature for
O K$\alpha$ (0.525 keV), Mg K$\alpha$ (1.25 keV), Al K$\alpha$
(1.49 keV), Mn K$\alpha$ (5.9 keV) and Mn K$\beta$ (6.4 keV).
}
\label{cal}
\end{figure}

Ground calibrations of the payload were done in two steps. In the first
step, we calibrated the camera package i.e. the H2RG detector, Cryo SIDECAR, 
and the CIB. In the second step, we did end-to-end calibration of the 
complete payload, i.e. the camera package, grating module, mirror module with
the flight electronics inside the rocket experiment section. The detector
was planned to be initially coooled to 130 K, and during the flight,
the temperature would drift up from 130 K. Therefore, we calibrated 
the camera package at multiple temperatures (from 125 k $-$ 170 K) and at
multiple line energies (0.525 keV $-$ 6.4 keV). Data were collected 
in 32 channel and processed 
by the same IDL pipeline for cleaning, event grading based on Swift XRT
grading scheme \citep{burrows05_swift}, and finally to 
generate the spectra. The top-left plot in Fig. \ref{cal} shows
the linearity of the detector at multiple temperatures for 
O K$\alpha$ (0.525 keV), Mg K$\alpha$ (1.25 keV), Al K$\alpha$
(1.49 keV), Mn K$\alpha$ (5.9 keV) and Mn K$\beta$ (6.4 keV) lines. 
We estimated the gain and offset for each of these temperatures. 
The top-right plot
in Fig. \ref{cal} shows the variation of gain as function of temperature. 
Knowledge of variation of gain with temperature is important in order
to correct the flight data for the drift in temperature.
The bottom-left plot shows the read noise of the camera as a function
of temperature. At 125 K, the read noise is around 6.5 e- and increases 
steadily till 160 K and then starts leveling out. Dependence of
energy resolution on the temperature for different energy lines 
is shown in the bottom-right plot. 
From these results, it is evident that there is no significant change
in gain, resolution and read noise in 125 K $-$ 135 K region. At temperatures beyond 150 K for the detector and 185 K for the SIDECAR, the detector noise and energy resolution begin to increase. All of these measurements were repeated multiple times.    

\begin{figure} [ht]
   \begin{center}
   \includegraphics[scale=0.09]{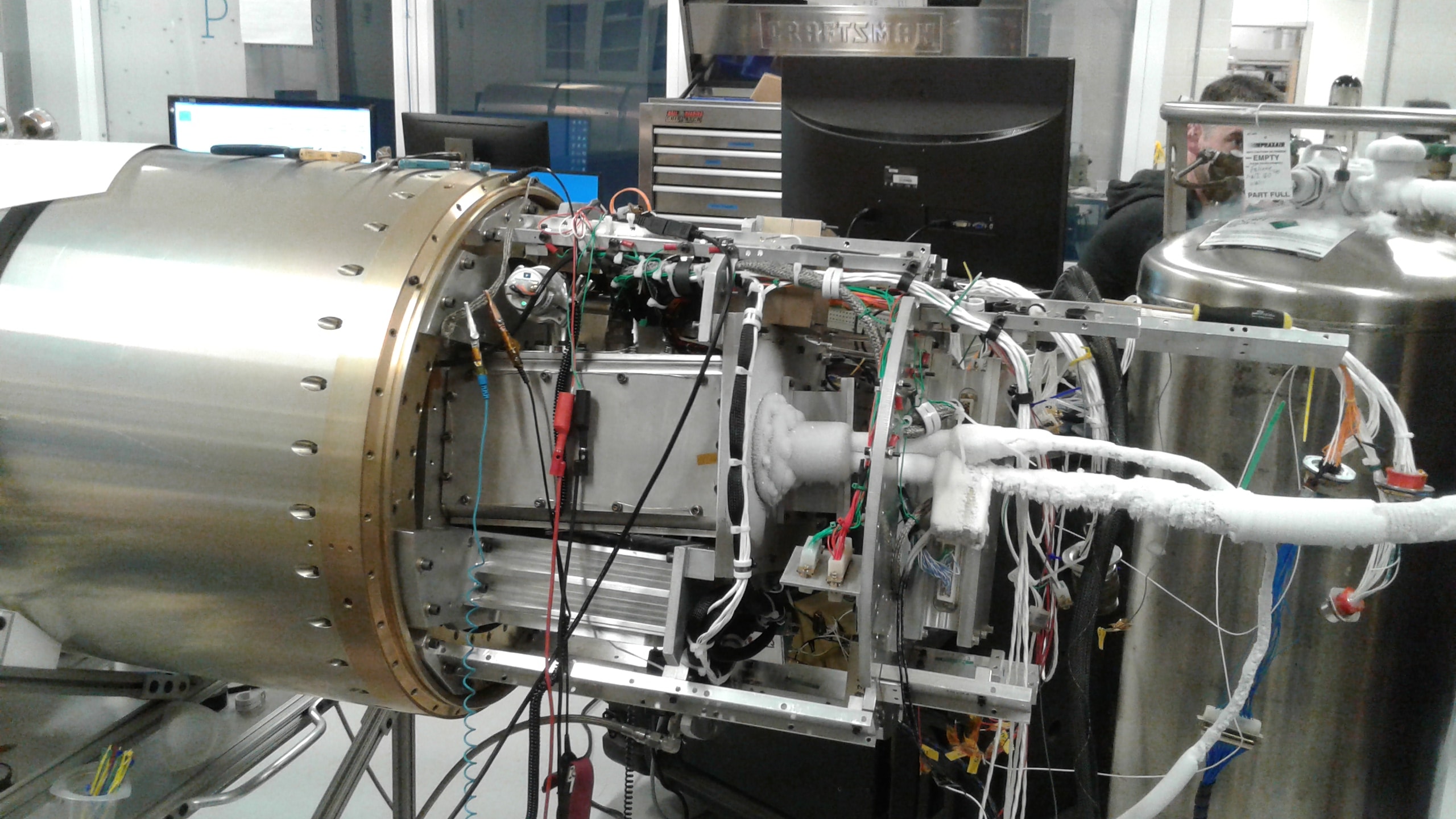}
 \end{center}
   \caption
{End-to-end calibration of the rocket payload. The grating and the 
mirror module is inside the rocket
chamber. The camera package chamber can be seen in the picture. The 
liquid Nitrogen (LN$_2$) cylinder (shown in the image) supplies LN$_2$ to cool down the detector and the 
Cryo-SIDECAR during calibration. An X-ray source (Manson source)
is connected to the other end of the rocket chamber to generate
X-rays. 
}
\label{rocket}
\end{figure}

\begin{figure} [ht]
   \begin{center}
   \includegraphics[scale=0.5]{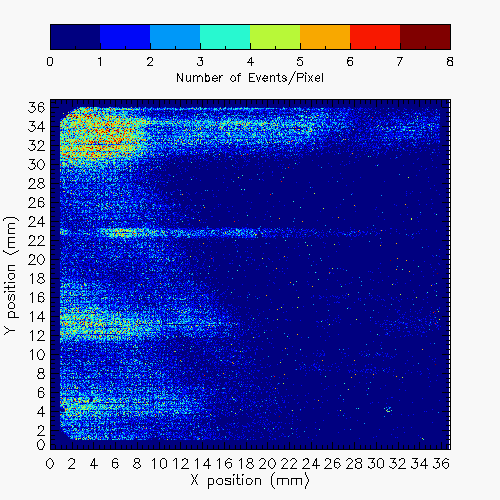}\\
\includegraphics[scale=0.25]{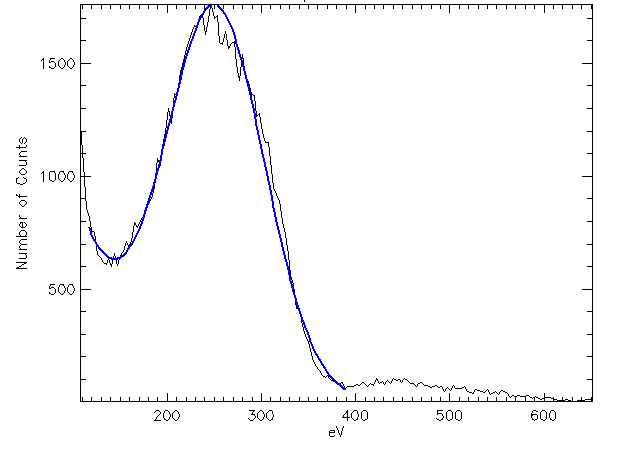}
\includegraphics[scale=0.26]{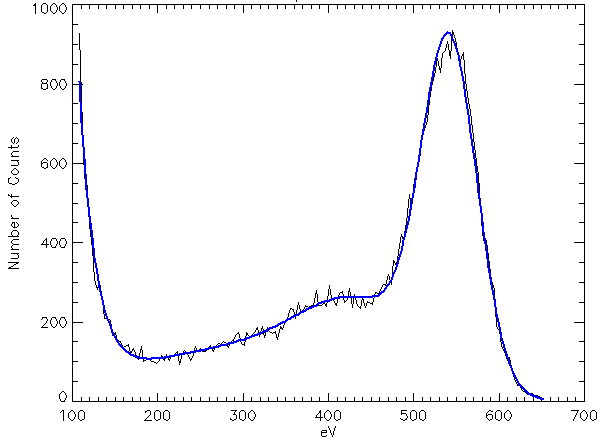}
\includegraphics[scale=0.25]{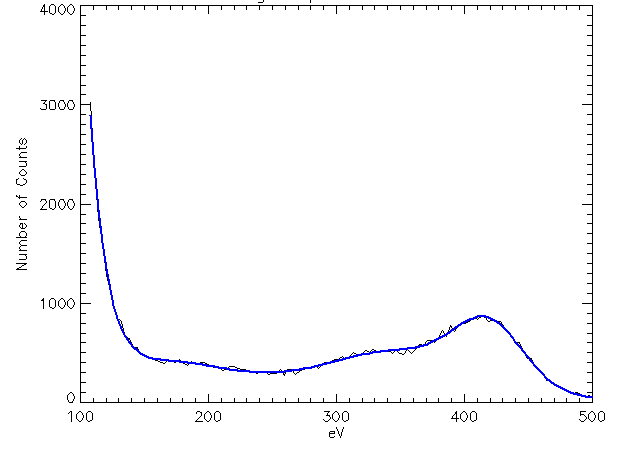}
 \end{center}
   \caption
{Top: 2D histogram of the X-ray events in the detector plane. 
The number of events per pixel is shown in colors. The grating 
module dispersed the X-ray photons of different energies to different 
regions of the detector where the top, middle and the bottom regions
correspond to C K$\alpha$ (0.27 keV), O K$\alpha$ (0.525 keV), N K$\alpha$
(0.39 keV) respectively.
Bottom (left to right): The corresponding spectra of C K$\alpha$ (0.27 keV), O K$\alpha$ (0.525 keV), and N K$\alpha$ (0.39 keV) lines. The blue solid lines
are the fit to the spectra.
}
\label{end_cal}
\end{figure}

In the second step, we calibrated the camera package inside the rocket 
experiment section with the grating module and the mirror module
as shown in Fig. \ref{rocket}. 
A Manson X-ray source was used 
to generate X-rays from the other end of the rocket section. The Manson
source uses a number of target anodes (Carbon, Magnesium, 
Aluminum, Copper etc.) where
electrons from hot filament are accelerated towards the target 
by a strong electric field. The electrons generate bremsstrahlung radiation
plus the K$\alpha$ lines from the anodes. Fig \ref{end_cal} (top) 
shows the histogram of events in the detector plane with different colors
indicating number of events per pixel.    
The grating module disperse the spectral lines in different regions 
in the detector plane. Three different lines are clearly visible in the image
where the line at the top region corresponds to Carbon K$\alpha$ line, 
the middle region corresponds to Oxygen K$\alpha$ and the region 
at the bottom corresponds to Nitrogen K$\alpha$ line. The spectra  obtained from these regions are shown in the bottom panel of the figure.

\section{recent developments in X-ray HCDs}\label{other}

\subsection{Small Pixel HCDs}
\begin{figure} [ht]
   \begin{center}
   \includegraphics[scale=0.052]{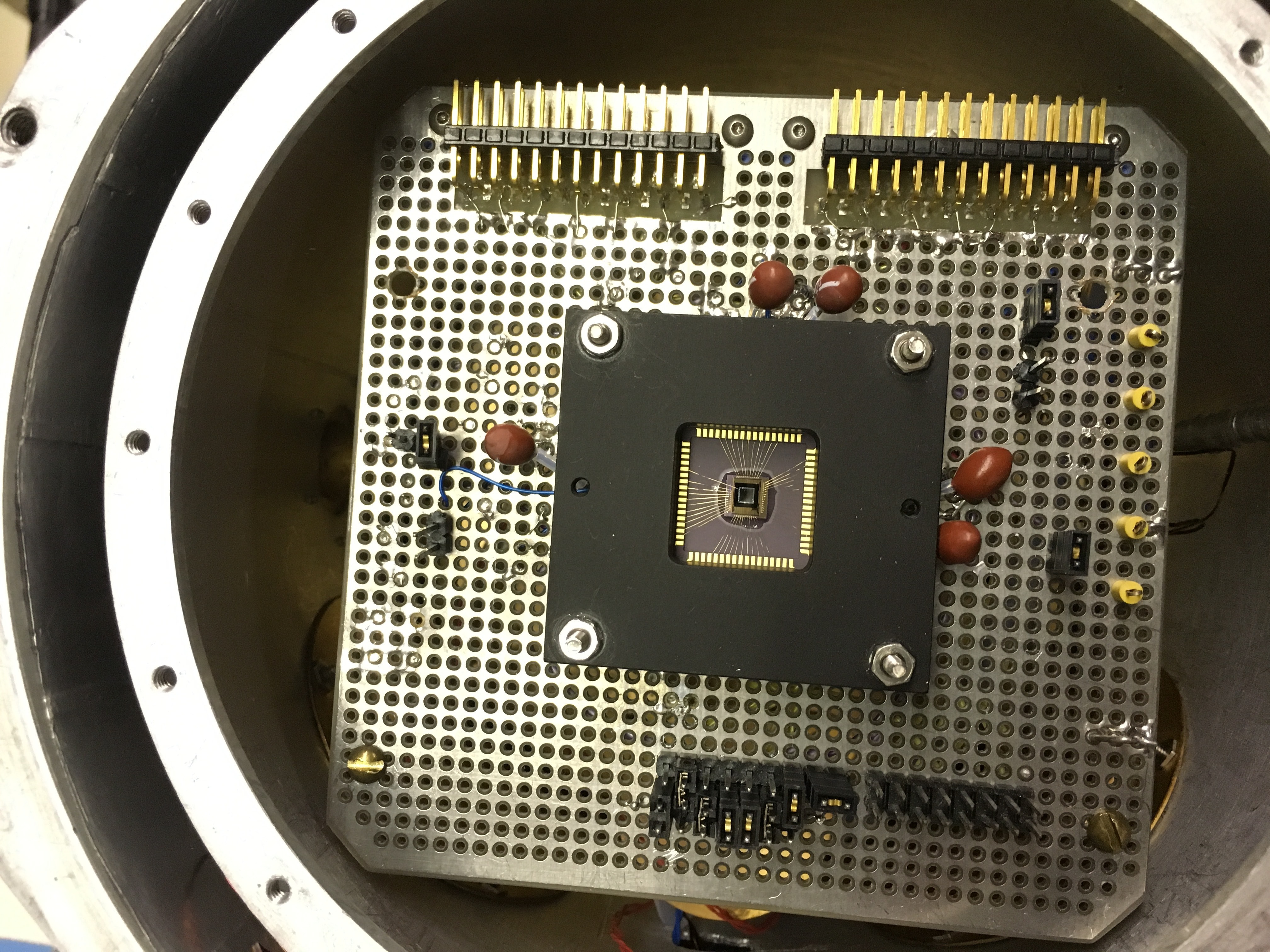}
\includegraphics[scale=0.236]{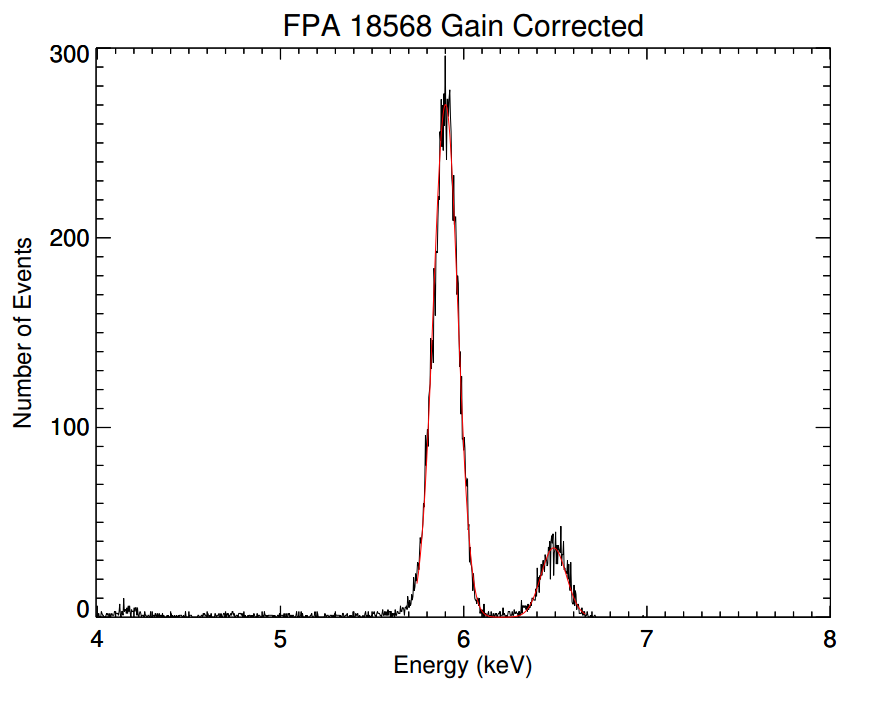}
 \end{center}
   \caption
{Left: Small pixel FPA 18567 seated in breadboard inside testing dewar. The breadboard measures 10 cm $\times$ 10 cm. Right: Gain corrected Fe$^{55}$ X-ray spectra for FPA 18568, showing the Mn K$\alpha$ and Mn K$\beta$ lines at 5.9 keV and 6.4 keV. See text for details.
}
\label{small_hcd}
\end{figure}

The small pixel HCDs are new prototype HCDs developed in collaboration between PSU and TIS. The detectors
were designed to satisfy the requirements of future fine angular resolution X-ray missions (e.g. Lynx): small pixel size, fast readout, and $<$ 4 e$^-$ read noise. The 128 x 128 prototype arrays have 12.5 $\mu$m pixel pitch with 100 $\mu$m fully depleted depth. The detectors use CTIA amplifier in each pixel, which unlike the source follower amplifier in
the HAWAII HCDs, holds the gate voltage constant during integration and eliminates inter-pixel capacitive crosstalk. The arrays also have the capability to perform in-pixel correlated double sampling (CDS) by subtracting the variable baseline voltage associated with a reset and thus remove reset KTC noise. Fig. \ref{small_hcd} (left) shows the picture of an engineering grade small pixel HCD test device.

Recent work has involved characterization of the small pixel detectors, including estimation of the pixel-to-pixel gain variation. The gain variation was calculated to be $\sim 1.1$\%, and gain corrected energy spectra were obtained. Fig. \ref{small_hcd} (right) shows the Mn K$\alpha$ and Mn K$\beta$ spectrum after 
correcting for the gain variation --- the measured energy resolution at
5.9 keV is $\sim$150 eV ($\sim2.7$\%) --- while the read noise is measured to be as low as 5.54 e$^-$. 
These are the best ever energy resolution and read noise measurements for any X-ray HCD detector reported to-date. 
Details of the experiment and these characterizations can be found in \citep{hull18_small_pixel}.
Currently, we are scaling this test design up to a large format detector with on-chip digitization, while attempting to further reduce read noise through improved component tolerances.

\subsection{Speedster-EXD} 
\begin{figure} [ht]
   \begin{center}
   \includegraphics[scale=0.09]{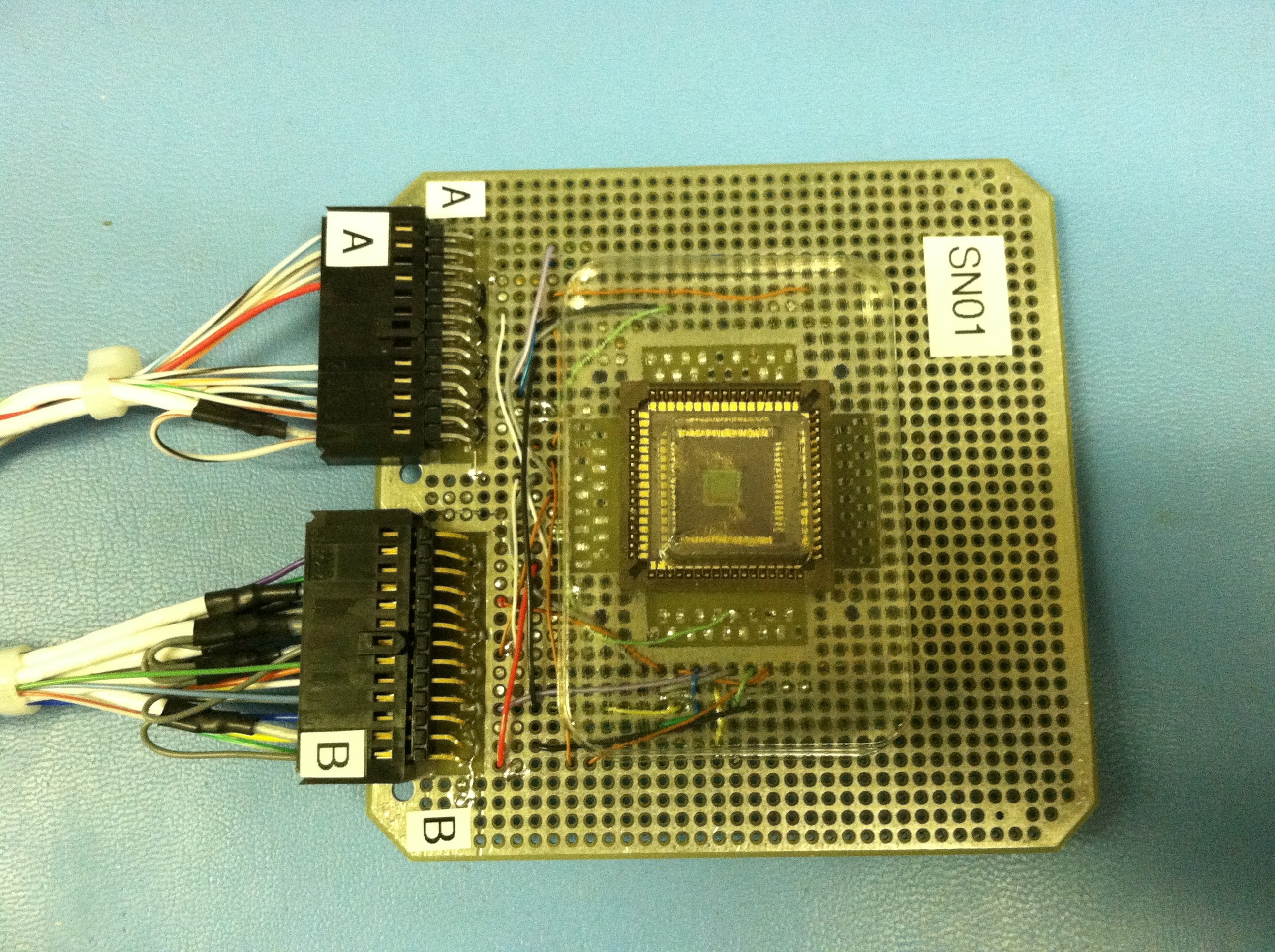}
 \end{center}
   \caption
{Photo of a prototype of Speedster-EXD detector. There are total 
64 $\times$ 64 pixels with 40 $\mu$m pixel size. To
block the optical light, there is a 50 nm Al layer directly deposited on the Si layer.
}
\label{speedster}
\end{figure}

\begin{figure} [ht]
   \begin{center}
   \includegraphics[scale=0.32]{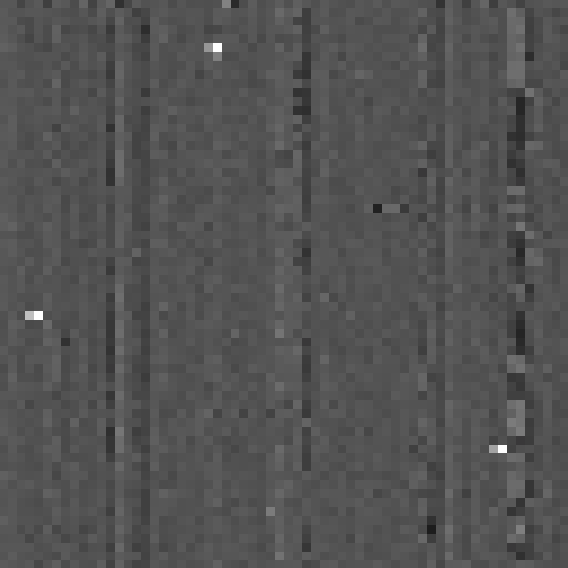}
\includegraphics[scale=0.265]{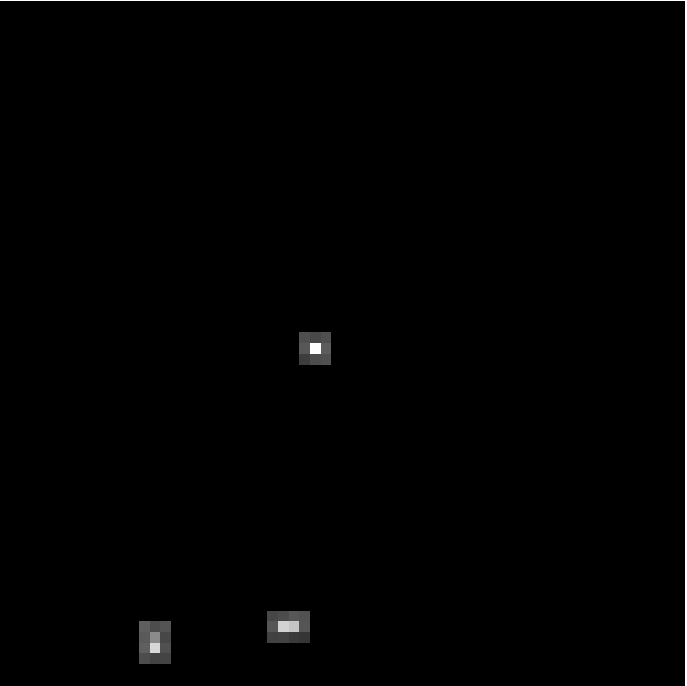}
 \end{center}
   \caption
{
Left: Full Frame Readout Mode. The comparator is set below the noise floor and thus every pixel in the
array is read out. Three multi-pixel Fe$^{55}$ events can be seen, in addition to all background pixels. Right : 3 $\times$ 3 Sparse
Readout Mode. The comparator is set above the noise floor, allowing read out of only X-ray events and their neighboring
pixels. Two multi-pixel and one single pixel Fe$^{55}$ X-ray events 
can be seen, with their 3 $\times$ 3 regions.
}
\label{speedster_readout}
\end{figure}

The Speedster-EXD is an advanced design utilizing a CTIA, in-pixel CDS, and event recognition circuitry to
provide event-driven sparse readout. Event-driven readout is accomplished using a comparator that samples
the CDS output and triggers on X-ray events. This provides timing resolution improvements by several orders
of magnitude and a significant improvement in high count-rate applications. 
Fig. \ref{speedster} shows an image of a Speedster detector 
with 64 $\times$ 64 
pixel array with 40 $\mu$m pixel size and 100 $\mu$m depletion depth. 
Fig. \ref{speedster_readout} shows the comparison of full frame 
readout mode (where the comparator threshold is set below the noise floor) 
and sparse readout mode (comparator threshold $>$ noise floor). 
In the sparse readout mode, the 3 $\times$ 3 region of
pixels surrounding the pixels above comparator threshold is read out. The Speedster-EXD prototype devices
have been characterized in terms of read noise, energy resolution, IPC, dark current, pixel-to-pixel gain variation.
Energy resolution was found to be around 3.5 \% at 5.9 keV. IPC was found to be negligible. More details on
the experiment set-up and characterization can be found in \citep{griffith16}.
A larger version of Speedster detectors is currently being developed. The new detectors will be 2.2 cm $\times$ 2.2
cm in size, with 550 $\times$ 550 pixel array (40 $\mu$m pixel size). The input cell architecture also supports multiple
readout nodes in order to maintain the high readout speed. The detectors also include on-chip analog-to-digital
conversion.

\section{Other ongoing activities with HCDs}
\subsection{Sub-pixel resolution in HCDs}
It is possible to exploit the charge sharing effect in pixelated detectors to find the exact location of the incident X-ray photons, which in turn can increase the scientific yield of a single observation by overcoming limitations of pixel size by more thoroughly sampling the PSF of X-ray telescopes. Recently, we performed a ``mesh experiment'' to experimentally determine the charge cloud shape for a custom H2RG with 36 $\mu$m pixels at 1.5 keV. For more details, refer to \citep{Bray2018JATIS}. Once the charge cloud shape is known, X-ray landing location can be determined with subpixel resolution. We found that X-ray landing location can be accurately determined in directions that exhibit charge sharing between pixels. For a 2-pixel event, the positional resolution can be obtained with an accuracy of $\sim$0.5 $\mu$m in the direction of charge split and $\sim$7 $\mu$m in the other direction at 68\% confidence level. For a 3-pixel event, the position resolution improves to $\sim$0.5 $\mu$m (68\% confidence) in both directions. 

Based on these experimental results, we have developed a model to predict the charge sharing and subpixel resolution as a function of X-ray landing position and applied it to a prototype 12.5 $\mu$m pitch small-pixel HCD that would be suitable for the Lynx X-ray surveyor \citep{Bray2018_SPIE_subpixel}. When operated with a substrate voltage of 80V, we determined a spatial resolution of 0.2 $-$ 1.2 $\mu$m (68\% confidence) for all events. When operated at 15V, the charge cloud becomes significantly larger, and we determine a spatial resolution of 0.4 $-$ 0.6 $\mu$m (68\% confidence) for all events. A summary of subpixel resolution results is shown in Table \ref{table:subpixel resolution summary}.
\begin{table}[ht]
\centering
\begin{tabular}{l|cc|cc|cc|}
\cline{2-7}
 & \multicolumn{6}{c|}{68$\%$ Confidence region half-width ($\mu$m)} \\ \cline{2-7} 
 & \multicolumn{2}{c|}{\textbf{H2RG}} & \multicolumn{2}{c|}{\textbf{\begin{tabular}[c]{@{}c@{}}Small pixel HCD\\  (V$_\textnormal{{sub}}$=80V)\end{tabular}}} & \multicolumn{2}{c|}{\textbf{\begin{tabular}[c]{@{}c@{}}Small pixel HCD\\  (V$_\textnormal{{sub}}$=15V)\end{tabular}}} \\ \hline
\multicolumn{1}{|c|}{\textbf{Pixel Pitch:}} & \multicolumn{2}{c|}{36 $\mu$m} & \multicolumn{2}{c|}{12.5 $\mu$m} & \multicolumn{2}{c|}{12.5 $\mu$m} \\ \hline
\multicolumn{1}{|c|}{\textbf{\begin{tabular}[c]{@{}c@{}}X-ray landing location\\ within pixel\end{tabular}}} & x & y & x & y & x & y \\ \hline
\multicolumn{1}{|l|}{Center} & 7.1 & 7.1 & 1.2 & 1.2 & 0.6 & 0.6 \\
\multicolumn{1}{|l|}{Right} & 0.4 & 7.1 & 0.2 & 1.2 & 0.4 & 0.4 \\
\multicolumn{1}{|l|}{Bottom} & 7.1 & 0.4 & 1.2 & 0.2 & 0.4 & 0.4 \\
\multicolumn{1}{|l|}{Bottom-right} & 0.4 & 0.4 & 0.2 & 0.2 & 0.4 & 0.4 \\ \hline
\\
\end{tabular}

\caption{A quantitative summary of the subpixel resolutions that can be achieved with the H2RG and small pixel HCD. X-ray landing location can be constrained very well in directions that exhibit charge sharing with neighboring pixels.}
\label{table:subpixel resolution summary}
\end{table}

\subsection{Radiation hardness of HCDs}
Due to the short distance that collected charge must be transferred through the detector, HCDs are inherently radiation hard relative to CCDs. Once in orbit, the primary cause of performance degradation is displacement damage caused by high energy protons trapped in the radiation belts. In order to investigate the effect of irradiation in HCDs, we irradiated an H1RG HCD with 8 MeV protons, up to a total dose of 3 krad(Si) (4.5$\times$10$^9$ protons/cm$^2$) \citep{Bray2018_SPIE_radiation}. The experiment was conducted at the Edwards Accelerator Lab, which is operated by the Physics Department at Ohio University. After irradiation, we performed detailed characterization of the detector to investigate the effects of irradiation on read noise, dark current, gain, and energy resolution. After 3 months, detector characteristics appear to be returning to the 
pre-irradiation level. We plan to continue this collaboration in the coming months by returning to Ohio University to carry out further irradiations of
higher doses.

\subsection{Wide field imaging with HCDs}
\begin{figure} [ht]
   \begin{center}
   \includegraphics[scale=0.6]{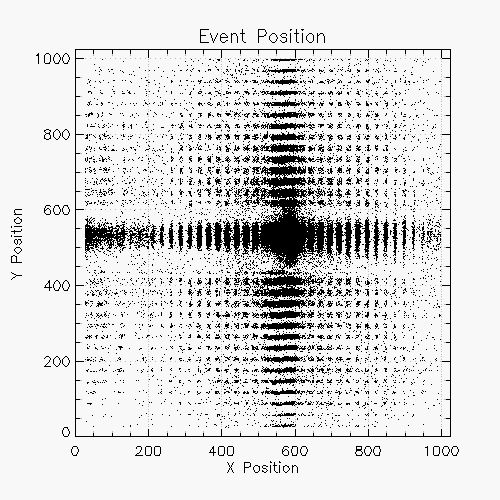}
 \end{center}
   \caption
{Point Spread Funtion (PSF) of the 2D lobster at 1.49 keV. The 
angular resolution is estimated to be $\sim$6 arc-min.
}
\label{psf}
\end{figure}

We are exploring the possibility to utilize HCD detectors as a wide field
imager to study GRBs and X-ray black hole transients for future 
CubeSat missions. The low power requirement of the HCDs and their fast 
readout make them suitable for the study of transients in CubeSat missions. 
In this context,
we have initiated characterization of an H1RG detector and a 2D
Schmidt Lobster optic \citep{hudec10_lobster} in 
collaboration with Czech Republic
Technical University (CTU). We utilize the 50 meter beam line at PSU
to characterize the mirror at multiple energies and multiple 
off-axis angles. The optic FOV is 5$^\circ$ $\times$ 5$^\circ$ and
30 cm focal length (See \citep{chattopadhyay18_lobster}
for more details). The left plot in Fig. \ref{psf} shows the obtained 
Point Spread Function (PSF) of the Lobster at 1.5 keV. 
The angular
resolution is found to be $\sim$6 arc-min. The details of the 
experiment set up and results will be reported elsewhere. 

Recently, we submitted a 6U class CubeSat proposal to NASA for a 
wide field imaging instrument consisting of an array of X-ray hybrid CMOS detectors 
at the focal plane of a coded aperture mask (BlackCAT; the Black hole Coded Aperture Telescope). 
The FOV of the instrument is 
$\sim$60$^{\circ}$ $\times$ 60$^{\circ}$ and sensitive in 0.5 $-$ 20 keV. 
The primary science goals of the mission are to detect and locate 
high redshift 
GRBs and the electro-magnetic counterparts to the gravity waves with 
sub-arcmin accuracy, while monitoring the X-ray sky to detect and study black hole transients. 
For more details, on the 
conceptual design of the instrument and science goals, see \citep{chattopadhyay18_blackcat}.

\section{Summary}
The PSU X-ray detector lab is involved in various activities relating to 
hybrid CMOS detectors and their continued technological development. 
We demonstrated significant improvement in energy resolution and 
read noise for an H2RG when the SIDECAR ASIC is cooled down below 180 K. 
The same H2RG detector and the 
cryo-SIDECAR was flown successfully in NASA's first Water Recovery X-ray Rocket flight on April 4$^{th}$, 2018 from Kwajalein Atoll. The success of the mission 
raised the TRL of X-ray hybrid CMOS detectors to TRL-9, demonstrating their suitability for future space missions. The TRL of the Camera 
Interface Board (CIB) which was developed at PSU to interface the SIDECAR 
and the spacecraft has also been raised to TRL-9. 
Small pixel HCDs with 12.5 $\mu$m pixel pitch, 
in-pixel CDS, and CTIA amplifiers have also been characterized, with 
$\sim$5.5 e- read noise measured. 
After measuring the pixel-to-pixel gain variation, the resultant spectra for the small pixel HCDs provides energy resolution of $\sim$2.7 \% at 5.9 keV, which is very close to the Fano limit and is the best ever reported for X-ray 
hybrid CMOS detectors. Speedster-EXD prototypes are shown to 
improve the readout speed of HCDs by a several orders of magnitude
and currently PSU and TIS are working on the  
development of a large array Speedster-EXD device. By performing a mesh experiment, we also demonstrated the 
feasibility of achieving sub-pixel resolution with HCDs. Other efforts in our lab include investigation 
of proton irradiation effects on HCDs, thereby demonstrating the radiation 
hardness of HCDs. We are also planning to utilize X-ray HCDs as a wide 
field imager
for future CubeSat missions to detect and study gravity wave event counterparts, high redshift GRBs, and black hole transients.   

\acknowledgments % equivalent to \section*{ACKNOWLEDGMENTS}       
We gratefully acknowledge Teledyne Imaging Sensors, particularly Vincent 
Douence, Mihail Milkov, Steven Chen, Mark Farris, and Yibin Bai for their very useful troubleshooting assistance and 
their excellent design work. This work was supported by NASA grants 
NNX13AE57G, NNX17AE35G, NNX14AH68G, NNX17AD87G, NNX16AE27G, and 80NSSC18K0147.

% References
%\bibliographystyle{spiebib}
%\bibliography{report}

\end{document}